# Cache Discovery Policies of MANET

Amer O. Abu Salem

Department of CS
Zarqa University
Zarqa, Jordan

Tareq Alhmiedat

Department of IT
Tabuk University
Tabuk, KSA

Ghassan Samara

Department of CS
Zarqa University
Zarqa, Jordan

Abstract—In situations where establishing a network infrastructure is impossible, Ad-hoc networks are considered particularly important. Most of the previous research in Ad-hoc networks concentrated on the development and enhancement of dynamic routing protocols, which could efficiently discover routes between two communicating nodes. Although routing strategies is an important topic in MANETs, other topics such as data access are also crucial since the final goal of using Ad-hoc networks is to provide data access to mobile nodes. One of the most attractive techniques used to improve the data access performance in MANET environment is cooperative caching; which means multiple caching nodes share and cooperatively manage the cached contents. It is lead the research to important questions, what data should be cached, where, when, and how? A cooperative caching addressed into two basic issues: cache discovery and cache management, in other words, how to find requested data efficiently and how to manage an individual cache to improve the overall capacity of a cooperated cache. In this paper we have made a review of the existing cache discovery algorithms to address four stages after application request and before server response, using an historical file to record the previous data requests, and proposed cluster architecture with data cluster head election to store efficient information for future use and reducing the cost of flooding. In addition, this paper suggests some alternative techniques for cache discovery. Finally, the paper concludes with a discussion on future research directions.

## I. INTRODUCTION

A Mobile Ad-hoc Network (MANET) is the active research topic in wireless communications. It is a collection of two or more nodes in which the communication links are wireless; without the service of any fixed infrastructure or centralized administrator. The network is Ad-hoc because each node is able to receive and forward data to other nodes, and so the judgment of which nodes forward data is made dynamically based on the network connectivity. The advantage of this type of network is that it does not require any kind of infrastructure, like a base station in cellular network. This paper has been structured as follows. Section 1 gives an overview of mobile Ad-hoc networks, cooperative caching strategy used and the problem statement. In section 2 we summarize the related works in caching of MANET . Section 3 describes the hybrid proposed approach which used to enhance the performance of MANET. Section 4 summarizes the results the evaluation of that case scenario and the conclusions.

### 1.1 MOBILE AD-HOC NETWORKS (MANETs)

A MANET is collection of wireless mobile nodes which forms a network without the assistance of any predefined network infrastructure [1]. The best features of MANETs include ease of deployment, speed of deployment, and independence of infrastructure. Therefore Ad-Hoc networks are best suitable for an environment, which is not able to provide any kind of infrastructure [2]. In MANETs, heterogeneous nodes all the time move in the environment, hence always changing the network's topology. Ad-hoc networks are decentralized without any controlling entity. Frequently nodes cannot communicate directly with each other; therefore, routes between nodes may need to pass over several hops in order to reach a destination.

To date, there exist several proposed various routing algorithms for Ad-hoc networks, each with their special advantages and disadvantages. Researchers traditionally sort out these protocols as proactive protocols, reactive protocols, or hybrid of them, based on the algorithms that find new routes or update existing ones [3].

MANETs are restricted by discontinuous network connections, limited power supplies, and computing resources etc. These limitations raise several challenges for data access applications with the respects of data availability and access efficiency. The most common data access applications of Ad-hoc networks are: content sharing, a user shares some locally stored information, such as music, video, and document files, with other nodes on the network and on the other side, the user may search for and discover wanted content at other nodes for downloading from these nodes; - instant messaging, a user keeps a contact list of other users, when any of these users are online, that user can chat with them; - distributed processing, a node plays the role of a server by sending raw data to process it to perform large-scale computations [4].

In critical environments, like battlefield, disaster rescue, earthquake recovery, and exploration of an area, MANETs are greatly applied [5]. However, the success of wireless communications has resulted in MANET use in commercial applications, such as conferences or vehicular Ad-hoc networks. In these new scenarios, users need access to external networks, particularly via the Internet.

The multi-hop communication admits the remote information servers and this leads to longer query latency and high power consumption. When several clients repeatedly access the information server, huge load is caused. Then the server response may be reduced due to huge load. Data caching is capable to overcome these restrictions that results query delay and bandwidth can be reduced efficiently [1].

*1.2. CACHING*

The Ad-Hoc network users deal with a collection of data and programs that are specific to the application and must be shared by all the nodes. These data must be distributed among the mobile devices. When a mobile node needs a copy of it, the data should be transmitted over radio links that are very expensive in power consumption, thus data movements must be limited and fully considered.

In wireless Ad-hoc networks, the data caching strategy becomes a problem. Caching data at nodes helps a wireless network system to run faster, reduce the cost of bandwidth and avoid overload at nodes with more efficiently. After it has been decided to cache, the next issue is what to cache and where [6].

Data caching and replication help to increase the availability of data, a necessity in the wireless network environment where the system is dynamic and unstable, which makes nodes enter and leave the system any time. In addition, the caching can help enhance the performance of request processing because it can assist reduction of the routing path of a data request because the requested data may be situated near the requested node. The critical issue in caching is, which data should be cached and when the cache is full, which data in the cache should be removed. So, a cooperative caching strategy needs to be specified: How are users' requests processed using the cooperative caching system? That is, once a caching node receives a data request, how does it proceed to resolve the request? And how does a caching node manage the cached data on behalf of the cooperative caching system?

*1.3. COOPERATIVE CACHING*

Often, to coordinate more cache in a network and share resources to serve other applications, it is also known as cooperative caching. If a node does not have the requested data item in its local cache, it can send the data request to a close cooperating cache that can serve the requested data faster than the original data server.

Battery power limitation, insufficient bandwidth, average latency, data availability and accessibility in MANETs are quite difficult due to the mobility of the nodes. Cooperative caching can improve this issue effectively. Cache cooperation can be between caches within a system or may even be across systems, such as cooperation between different Internet service providers. Several mechanisms and protocols have been proposed to determine how the cache should communicate and determine which items to cache.

As mentioned before, MANET has special properties that require special features in the design of cooperative caching schemes for its environment. The importance of the cooperation is a caching node serves not only the data requests from its applications, but also to data requests from other nodes applications, and a caching node not only saves data on behalf of their own requests, but also on behalf of the other nodes' requests [7].

In that order, a cooperative caching strategies need to define two main problems: once a node receives a data request, how does it proceed to resolve that data request? And how does a caching node manage the cached data on behalf of the cooperative caching scheme?

*1.4. MANET CACHING FEATURES*

MANET is different from architecture based wireless network that there are no dedicated network infrastructure devices in it. Because of the limited radio range of wireless devices, the route from one device to another may demand multiple hops. In some scenarios, to communicate with the outside network, only some devices that have network connections with the outside base stations or satellite system can serve as the gateways for the Ad-hoc network. Researchers would refer to such Ad-hoc networks as weakly connected Ad-hoc networks, since there is some infrastructure support, but not as very much like the mobile nodes in the infrastructure based networks.

In general, the MANET environment also has the two characteristics of wireless computing environments; weak connectivity and resource constraints. Therefore, there is need to concentrate on the problems of bandwidth efficiency and data availability.

- Compared to wired connections, the wireless connections of the MANET are of low bandwidth and subject to frequent disconnections for several reasons, leading to weakly connected mobile nodes. This feature produces a new critical problem for cache management: how to achieve high data availability in mobile Ad-hoc environments where many disconnections may happen since the mobile nodes and server may be weakly connected? The mobile nodes can frequently be disconnected from the data servers either involuntarily if wireless connectivity may not be available in the areas, or voluntarily if the user shuts off the wireless network interface to save up battery.

- A second feature is resource constraints of the nodes, because typical nodes in MANET environments have limited power and processing resources. This feature extends to another critical problem for cache management in MANET environments, that is to say, how to minimize power and bandwidth overhead for cache management? Many schemes to solve these problems have been proposed.

## II. Problem Statement & Proposed Solution

As mentioned above, data access applications in MAENTs can experience low data availability and high latency, energy, and bandwidth costs due to multi-hop communication and dynamic network links. The cooperative caching is a solution to enhance the system performance in data access application over wireless networks such as the mobile Ad-hoc network as a scheme to improve data access efficiency, because it actually increases the effective cache size for mobile nodes, i.e. mobile nodes can enable more efficient access to data which are stored in another node's cache.

The aim of this research is to develop a hybrid cooperative cache discovery strategy to increase the efficiency and assist the improvement of the on-demand data access ratio and to reduce the local cache miss ratio in mobile Ad-hoc networks, based on historical file record of the previous data requests to store efficient information for future use, as a second stage in discovery procedure after local cache miss and used a cluster architecture, so it proposed a strategy to choose a data cluster head, to reduce unnecessary queries flooding by assigning specific tasks to that data cluster head.

## III. Related Work

In MANET the Internet gateways are fixed nodes as defined in [7]. The Web caching process runs as follows: a node requests a data item to a data server; the request is routed from the source to the destination in Ad-hoc network using the routing protocol defined for this network. When the data server receives this request it replies sending the data item to the requested node. This client-server scheme is much related to the one employed in the Internet for the HTTP Web traffic. To develop three process the paper of González is comparison between three common caching discovery schemes [8].The paper of Gitzenis proposes a modeling framework to study the problem of power-controlled data prefetching[9]. And the paper [10] proposed prefetching techniques to increase a good trade-off between latency and power cost in a broadcasting environment. Narottam Chand et al [11] have propose a novel scheme, called zone cooperative (ZC) for caching in mobile ad hoc networks. In ZC scheme, one-hop neighbours of a mobile client form a cooperative cache zone. Hence the cost for communication with them is low both in terms of energy consumption and message exchange. Cache admission control and VALUE based replacement policy are developed as a part of cache management. It improves the data accessibility and reduces the local cache miss ratio. They also performed an analytical study of ZC based on data popularity, node density and transmission range. The network is divided into clusters and the selection of cluster head is done based on the power level and connectivity. The details of the cached data items in the cluster and its adjacent clusters are maintained in the local cache table (LCT) and the Global cache table (GCT) [12].

## IV. Cache Discovery Policies for Cooperative Caching in Mobile Ad-hoc Networks

Cooperative Caching reduces communication cost of the system, which consequently results in the reduction of bandwidth and power consumption. The proposed approach developed as a middleware layer between a routing protocol and an application layer protocol as a proxy. It helps the lower layer communication protocols for detecting an efficient route to the owners of the data item, which is requested by an application layer protocol. In order to achieve this, neighboring nodes cache different data items, which in turn will increase the data diversity in a region. As a result, more data items will be accessible in the network.Generally, an effective cooperative cache model for MANETs should address these issues: An efficient cache discovery algorithm to discover and return requested data items from the neighbors of the requesting node. And there should be cache management strategies, which consist of the cache admission control algorithm, the cache replacement algorithm and the cache consistency algorithm. Here, sharing and coordination is processed among the multiple nodes that reduce the bandwidth and power consumption[13].

This paper focuses on all issues that related to efficient cache discovery, to improve the data availability, meaning data is available in minimum time and by utilizing more than a few resources. As an initial research to come up with the hybrid cooperative cache discovery and management scheme based on a combination of cache maintenance algorithms and using the MANET philosophy at the end.

### 4.1 Non-Cooperative And Hop-By-Hop Caching Strategies

The traditional way of resolving an application's data request, called non-cooperative caching scheme (NC), is to first check the local cache. If the local cache has a valid copy of the requested data item, the cached copy is returned to the application. However, if the cache misses, the request is forwarded to the data server to retrieve the requested data item. This scheme works well as long as the connection to the server is a reliable high speed connection and not too expensive. But for MANETs, every hop is a risk because of signal interference and link breaks induced by nodes movement. To increase data availability and reduce the cost of time and power, Hop-by-hop cache resolution allows a node on the forwarding route to check and serve as a proxy for its received request. If a forwarding node has an unexpired copy of the requested data in its local cache, it can stop forwarding the data request and sends back the requested data to the requester [14]. But the cooperation is very limited by just using hop-by-hop; it is only between upstream and downstream nodes.Figure 1 illustrates a scenario where non-cooperative and hop-by-hop can be improved. In this example, at case 1, MN4 already has a map of the UK as a data item in its local cache, after it reached the server to get it by using two schemes. After that, at case 2 assuming MN2 issues a request for the data item, and using non-cooperative scheme MN2's request for the UK map should reach the data server to get the data item. This will be the case regardless of the fact that MN4 is in the forwarding route and has the requested data. Using the hop-by-hop scheme before MN3 relays the request, it checks if it has a valid copy of the UK map in its local cache. If so, MN3 stops forwarding the request and sends the copy back to MN2 otherwise relay the request to next hop in the forwarding route. So MN2's request for the UK map should be able to get its answer from MN4, way ahead of reaching the data server.

Hop-by-hop only enables upstream nodes to answer data requests for downstream nodes; i.e. in case 3, if any forwarding mobile node, MN6 in this example, moves out of MN5's range, the data requests from downstream nodes must wait for the recreation of a new route to the data server. During this waiting period, requests can be missed because of a full queue. Hop-by-hop will not be able to assist in the data request, even though they are immediate neighbors, because a downstream node can reuse the cached data in an upstream node only.

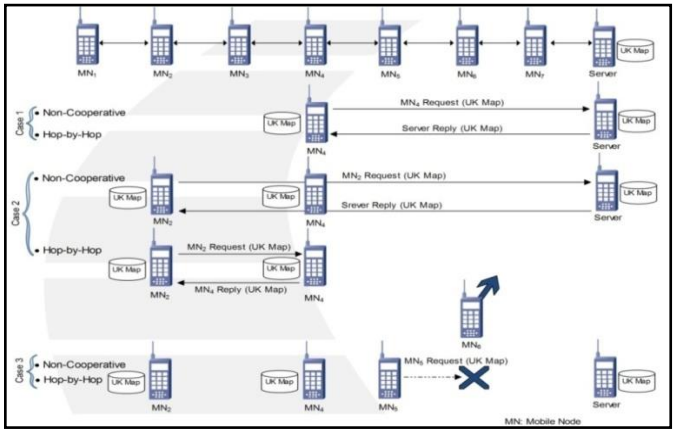

Figure 1. Non cooperative and hop-by-hop caching scenarios

To this end, this research studies how a mobile node can re-use any data item cached within its neighbors in a cooperation range of the node. The idea is that nodes in one cooperation range should have relatively faster and more reliable network connections between members; otherwise the cooperation will get significant communication overhead. In this example, MN5 will be able to get the requested data, the UK map, from a node in its cooperation range; MN4 in this case. In Ad-hoc networks, throughput and reliability of a network connection seriously depend on the number of hops covered by the connection [15]. As the number of hops in a route increase, the throughput and reliability decreases. The proposed approach combines the Hop-by-hop scheme and the cooperation range scheme and comes up with a hybrid approach to study the cache discovery problem in MANET. The emphasis of cache discovery in cooperative caching is to answer how nodes can help each other resolve data requests to improve the overall performance.

## V. Cache Discovery Scheme (Proposed Approach)

Because of the lack of a fixed communication infrastructure, one of the main issues in cooperative caching strategies is a cache discovery scheme that can efficiently discover and return requested data items from the local cache of neighbors' node to the requestor. This paper provides an attractive solution for cache discovery scheme. It proposes a new proactive approach for cooperative caching in MANETs; caching the information of previous requests in a previous request table (PreReq). It also proposes a new strategy to elect a powerful node as a data head cache state node (D-Clusterhead) that is responsible for maintaining the information of cluster cache state table of different nodes belonging to the cluster boundary. This will help to improve the data availability and overall performance of the network.

### 5.1. A Proactive Approach Contribution

With the absence of fixed infrastructure (e.g., no router, no access point, etc.), two nodes communicate with each other in a node to node approach. Two nodes communicate directly if they are in the transmission range of each other. Otherwise, nodes can communicate through a multi-hop route with the cooperation of other nodes. To find such a multi-hop route to other nodes, each MANET node widely uses flooding or broadcast techniques depending on hello messages. Flooding in MANETs is an important communication primitive and also attends as an essential technique for many Ad-hoc routing protocols [16], multicast schemes [17], or service discovery programs depend on huge flooding. Flooding is the mechanism by which a node, receiving flooded message m for the first time, rebroadcasts m once. With differences between flooding and broadcast, this is a transmission that is received by all nodes within transmission range of the broadcasting node. Flooding usually covers all the nodes in a network, but can also be limited to a set of nodes that is defined by a geographical area, called geocast flooding [18] or by the time-to-live (TTL) parameter of m. Thus, a node receiving the flooded message only rebroadcasts it if it is within the specified area or if the message's TTL is greater than 0.

The reactive approach is applied when a node is unsure if the requested data item is available in its range or not. In this instance, the node can reactively discover this by flooding the request inside its range. If a node receives a broadcasted request, it checks its own cache for the requested data item. If the node has a valid copy of the requested item, it replies the requested data item to the requesting node. Otherwise, it records the received request, which provides a location hint if this node requests the same item in future. Indeed, flooding leads to a large amount of redundant messages that consume scarce resources such as bandwidth and power and cause contention; collisions and thus additional packet loss [19]. Since mobile Ad-hoc network flooding generates a huge amount of traffic, it must be controlled as much as possible.

A proactive approach can be used to reduce the huge number of flooding and the discovery latency; a proposed approach depends on a historical-based discovery for data cache discovery. Each node maintains a historical file of previously received data requests called (PreReq). It is meant to avoid duplicated flooding for the same data item and to determine the closest data cache based on the previously received data requests and its status.

In this strategy, This PreReq table will contain n records where n is the number of the data items; there will be five entries related to each record:

- The first entry is the requested data ID.

- The second entry shows whether requested data di is locally cached at the node or not, if it is cached what is the size of data.

- The third entry shows which nodes have cached di related to the distance by hops.

- The forth entry maintained to access count that show how many times di is cached by neighbors node of node MNi after di is cached by node MNi.

- The final entry shows the time-to-live value that after how much time di is expired, this value is assigned by the data server.

The size of PreReq can be specified by the mobile operating system. If it is full, the newest request over-writes the oldest request record.

TABLE 1 - AN EXAMPLE OF PREREQ TABLE

| Req Data ID | Locally Cached : Size | Cached Nodes : Distance | Popularity | TTL |
|---|---|---|---|---|
| D8 | No | $MN_5$:8, $MN_7$:7 | 0 | NA |
| D6 | Yes : 50 | $MN_2$:2, $MN_6$:4, $MN_8$:7 | 3 | 01:00:00 |

The next sections illustrate how this strategy gives the node a new chance to check this table after its local cache misses and before broadcasting a request; to avoid the traffic overhead and to locate the closest data cache. Functionally, as following:

- A node checks PreReq table after its local cache misses and before the broadcast of a request.

- If matching entries are found, the node compares its network distances to these matching caches and the original data server, and then chooses the closest one to send the data item request.

- If the requested data is successfully returned, it means the PreReq table entry is still good.

- Otherwise if the data request fails, it means the cache recorded in the PreReq table entry does not have the requested data. Therefore the PreReq entry is moved out, and the next strategy is used to discover the data request.

In heterogeneous mobile Ad-hoc networks, some nodes may have more cache capacity and power than other nodes. In this instance, these powerful nodes will be more active in replying to other nodes' requests, as they can cache more data items in their local cache and remain active for a longer time.

The historical-based discovery can be extended by representing and recognizing these powerful nodes, such that next data requests can be redirected to the powerful nodes to get more helpful services. As explained previously, flooding can help to discover the closest cache around the requester node, as well as serve to recognize who has the data next time. However, network flooding generates a huge amount of traffic, and it should be used sparingly. The next section presents a detailed study for identification of these powerful nodes that will be more active in replying to other nodes' requests.

### 5.2. CLUSTERING APPROACH CONTRIBUTION

MANET is called a multi-hop network because each mobile node is connected to other nodes through intermediate nodes. Therefore the aggregation of nodes into groups provides a convenient framework for channel management, reducing the number of control message exchange, and providing for flexible move management. This logical node group is called a cluster and the process of building up a cluster is called cluster formation. This research doesn't focus on the problem of how to divide the network topology into optimal clusters, so it assumes that topology is divided into equal-area size of grids; however, any clustering algorithm can be applied.

D-Clusterhead Definition:The size of grid gets the maximum distance between two nodes inside a cluster. That means each node is at one hop distance from a neighbor inside the same cluster. Network area is assumed to be virtually extended such that boundary clusters also have the same size as other clusters. In this proposed clustering scheme, the grid size in the network topology is an important factor into network division. As all mobile nodes in one cluster can reach to every other cluster member in one hop communication, the value of g grid is derived as following:

While MNi and MNj in one hop communication, which should be the distance between them (dis) is less than or equal to transmission range (r), and the maximum distance (dis) between MNiand MNj is:

$$dis = \sqrt{g^2 + g^2} \Rightarrow dis = g\sqrt{2} \Rightarrow dis \le r \Rightarrow g\sqrt{2} \le r \Rightarrow g = \frac{r}{\sqrt{2}} \Rightarrow g = \frac{r}{\sqrt{2}}$$

So when, all cluster members can reach to one another in one hop communication.In heterogeneous mobile Ad-hoc networks, some nodes may be more powerful and have more capacity than other nodes in one cluster. In this case, these powerful nodes will be more active in answering other nodes' requests, as they can cache more data items in their local cache and be active for longer time. The proposed discovery scheme may be extended by discovering and managing these powerful nodes, such that future requests can be redirected to the powerful nodes to get better services. These powerful nodes are selected as cluster head for clusters in some of the clustering algorithms, which is use to receive and forward all the requests for a cluster members from/to nodes, which tends to make it a bottleneck and a point failure when the system has high network density[20].

To further improve the efficiency of the discovery scheme, this research proposes a new strategy to elect the most powerful node as a data clusterhead cache state node (D-Clusterhead) which is responsible for maintaining the information of cluster cache state table of different nodes belonging to the cluster boundary, and provide additional service during cache discovery and cache management . That state table is the list of data items along with their TTL stored in its cache, and there is a consistency scheme to update any modification at D-Clusterhead when a node cache or replace any data item in its local cache. Compared to clusterhead scheme, D-Clusterhead deals less workload, does not have to be as powerful as a clusterhead and depends on different parameters for the election algorithm.

D-Clusterhead Election Algorithm:To decide how well-suited a node is for election as a D-Clusterhead, take into account the D-Clusterhead election algorithm, that it is not periodic and is invoked as rarely as possible. This reduces system updates and hence computation and communication costs. So the weight function depends on the following parameters: cache size, distances, mobility, battery power and node popularity:

- Cache Size (CS): if the node has enough storage space, then caching is not a big problem, and it is able to cache and service better than other cluster members.

- Distances (D): a D-clusterhead is able to communicate better with its neighbours having closer distances from it within the transmission range.

- Mobility (M): is an important factor in deciding the D-Clusterhead. In order to avoid frequent D-Clusterhead changes, it is desirable to elect a D-Clusterhead that does not move very quickly. When the D-Clusterhead moves fast, the nodes may be detached from the D-Clusterhead.

- Battery Power (BP): the battery power can be efficiently used within certain transmission range, i.e., it will take less power for a node to communicate with other nodes if they are within close distance to each other. A D-Clusterhead consumes more battery power than an ordinary node since it has extra services.

- Popularity (P): If the node is often queried, then the node is popular, and it is a good idea to cache the resource.

And the next steps illustrate the procedure of D-Clusterhead election strategy:

- Step 1. For each node v, compute the free local cache capacity CSv

- Step 2. For each node v, compute the sum of the distances Dv, with all its neighbours (i.e., nodes within its transmission range and cluster), as

$$D_v = \sum_{v' \in N(v)} \{\text{dist}(v, v')\}.$$

- Step 3. Compute the running average of the speed for every node till current time T. This gives a measure of mobility and is denoted by Mv, as

$$M_v = \frac{1}{T} \sum_{t=1}^{T} \sqrt{(X_t - X_{t-1})^2 + (Y_t - Y_{t-1})^2},$$

Where $(X_t, Y_t)$ and $(X_{t-1}, Y_{t-1})$ are the coordinates of the node v at time t and t−1, respectively.

- Step 4. Calculate remain of BPv, during which a node v acts as a D-Clusterhead. BPv means how much battery power has been consumed, which is assumed a D-Clusterhead consumes more battery power than an ordinary node since it has extra.

- Step 5. For each node v, depending onPreReq compute the popularity of nodePv.

- Step 6. Calculate the combined weight Wv for each node v, where

$$W_v = 1/w_1 CS_v + w_2 D_v + w_3 M_v + w_4 BP_v + 1/w_5 P_v$$

Where w1,w2,w3,w4 and w5 are the weighing factors for the corresponding system parameters and w1+w2+w3+w4+ w5= 1.

- Step 7. Select that node with the smallest Wv as the D-Clusterhead. All the neighbours of the chosen D-Clusterhead are no longer allowed to participate in the election procedure.

- Step 8. Repeat steps 2–7 for the remaining nodes in other clusters not yet selected a D-Clusterhead.

The motivation of Dv is principally related to power consumption. It is proven that more power is required to communicate to a longer distance. The third component for Wv is due to mobility of the nodes. As mentioned before, a node with less mobility is always a better selection for a D-Clusterhead. The last component Pv, is measured as the total (cumulative) time a node acts as a D-Clusterhead. The model assumed the battery power of all nodes to be the same at the beginning. Also, take into consideration that the battery power will be more for nodes acting as D-Clusterheads. However, if the nodes have several battery powers to begin with, then it would be a more exact to measure the power currently available at the node. This depends on the node's initial power and the power spent based on actual network traffic and distance of the connections used to support it.

An Illustrative Example:This section demonstrates the proposed weighted data cache clusterhead election algorithm by example. All numeric values, as obtained from executing the algorithm on the 9 nodes in one cluster as shown in Figure 2, are tabulated in Table 2. The Figure shows the partitioning of a MANET into cluster grids depending on geographical area, configuration of the nodes in the network with individual node ids. Dotted squares represent the fixed cluster borders for each one. A node can connect its cluster members in one hop, and it shows one of these clusters. The arrows in the Figure represent the speed and direction of movement associated with every node. A longer arrow represents faster movement and a shorter arrow indicates slower movement.

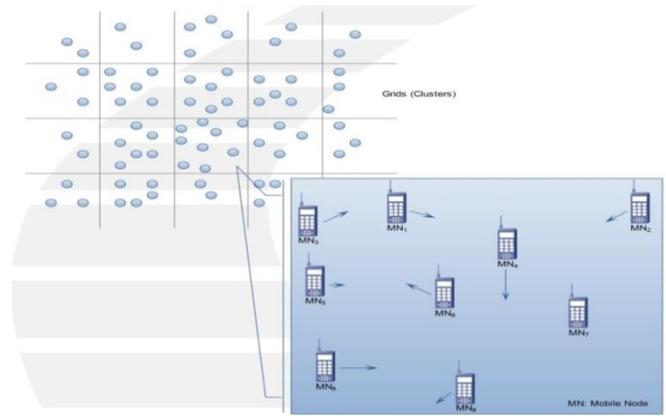

Figure 2. Nodes movement in one cluster

The values for CSv (step 1), are chosen randomly; CSv = 0 implies that a local cache of the node is full. Sum of the distances, Dv, for each node is calculated as step 2, where the unit distance has been chosen arbitrarily. The values for Mv (step 3), are chosen randomly also; Mv = 0 implies that a node does not move at all. Some arbitrary values have been chosen for BPv which represent the power battery of node. This corresponds to step 4 in the algorithm. The values for Pv (step 5), are chosen randomly.

After the values of all the components are identified, compute the weighted metric, Wv, for every node as proposed in step 6 in the algorithm. The weights considered are w1 = 0.5, w2 = 0.3, w3 = 0.1, w4=0.05 and w5 = 0.05. Note that these weighing factors are chosen arbitrarily such that w1 + w2 + w3 + w4 + w5 = 1. The contribution of the individual components can be tuned by choosing the appropriate combination of the weighing factors. At the end, in this example and depending on the proposed algorithm, MN6 elected as the optimal D-Clusterhead for that cluster.

| Node ID | CSv step 1 | Dv step 2 | Mv step 3 | BPv step 4 | Pv step 5 | Wv step 6 |
|---------|-----------|-----------|-----------|------------|-----------|-----------|
| MN₁ | 52 | 11 | 2 | 3 | 15 | 5.02 |
| MN₂ | 42 | 13 | 2 | 2 | 10 | 6.24 |
| MN₃ | 62 | 12 | 3 | 6 | 9 | 6.45 |
| MN₄ | 47 | 10 | 4 | 7 | 14 | 5.22 |
| MN₅ | 24 | 12 | 1 | 4 | 13 | 5.51 |
| MN₆ | 53 | 8 | 2 | 5 | 18 | 3.99 |
| MN₇ | 68 | 13 | 0 | 4 | 19 | 5.18 |
| MN₈ | 71 | 14 | 3 | 2 | 20 | 5.62 |
| MN₉ | 38 | 14 | 1 | 3 | 8 | 7.00 |

D-Clusterhead Maintenance:This paper proposes D-Clusterhead maintenance strategies in order to achieve better discovery approach in cooperative caching in MANETs.

1- When a node moves within the same cluster which cases no overhead in the proposed scheme.

2- When a non-D-Clusterhead node moves out of its current cluster :

- It informs its D-Clusterhead, to ask it to delete all entries related to this node.

- Then the D-Clusterhead will decide whether the data held by the leaving node can be cached by itself. These decisions will be based on following criteria: First the D-Clusterhead will check if the same data is cached by some node in their vicinity. If the entries matched then this data will not be cached by these nodes because this data is already available in the cluster. Otherwise this data will be cached by the D-Clusterhead. If the free space is not available in it, the data will be cached by any cluster member, else by using cache replacement algorithm some data is removed from the cache to save the new one.

3- When the node enters new cluster :

- Sends a broadcast request with its location coordinate (x,y) to one hop neighbours.

- The neighbors who belong to the same cluster will reply with cluster ID and D-Clusterhead ID.

-With the first positive reply, the node sends its cache information to the D-Clusterhead, and joins this cluster successfully.

- If there is no reply after threshold time, that means there is no neighbour in this cluster and that node becomes D-Clusterhead as it is the first node to enter this cluster.

4- When a D-Clusterhead node moves out of its current cluster to a new cluster:

- Before leave the current cluster, it selects the next most qualified node to be the new D-Clusterhead.

- Move all the cache state entries to the new D-Clusterhead.

- The new D-Clusterhead multicasts a change message to all cluster members with new D-Clusterhead ID.

- The old D-Clusterhead deletes all the cluster cache state entries and joins the new cluster as described above.

5- When a D-clusterhead does not respond to other cluster members:

- When a node communicates with the D-Clusterhead, it detects a specified period time, the node will renew this time before it expires. So if a node discovers that the D-Clusterhead is not available, it simply volunteers itself to initiate D-Clusterhead election by broadcasting an election message. Each node will reply with its evaluated value to the initiator in order for it to select an optimal node as a new D-Clusterhead, and then the new D-Clusterhead broadcasts its selection to all members of cluster. All member nodes update their D-Clusterhead ID and send cache information to the new D-Clusterhead.

6- When a non-D-Clusterhead does not respond to a D-Clusterhead:

- When a node does not renew the specified period time before it expires, the D-Clusterhead assumes that the node is failed; in this case D-Clusterhead removes the associated cache entry.

### 5.3. THE PROPOSED CACHE DISCOVERY ALGORITHM

These schemes described previously benefit different phases of cache discovery. Cache discovery scheme addresses not only how to resolve a data request with minimum cost of latency, power consumption and bandwidth, but also how to improve request success ratio. The importance of cache discovery in cooperative caching is to answer how mobile nodes can assist each other in resolving data requests to improve general network performance.

A cache discovery algorithm is required to determine if and where the requested data item has been cached when the requester does not have information of this destination. As shown in Figure 3, after a data request is initiated at a mobile node, the proposed approach categorize the discovery algorithm into four levels before arrival at the data server; it first looks for the data item in its own cache, if there is a local cache miss, the node looks in the PreReq table for the cached node ID that has a data item in its cache. If there is no ID available, it sends a request packet to its D-Clusterhead to verify if the data item is cached in other nodes within its home cluster. When a node receives the request packet and has the data item in its local cache i.e. cluster cache hit, it sends the requested data item or the cached node ID to the requester. In case of a cluster cache miss, the request is forwarded to the neighbor along the routing path to access the data server, and before forwarding a request, every node along the routing path searches for the item in its local cache or home cluster as described above. If the data item is not found on the cluster along the routing path i.e. remote cache miss, the request at the end reaches the data server. It can then finally return with the actual data item to the requester.

In the hybrid discovery scheme, the order of the basic strategies is based on feasibility consideration. For example, if the hop-by-hop scheme is set before the home cluster discovery, it means that the data request is first sent to the data server and if none of the forwarding nodes or the server can reply the data request, the request will be forwarded to the D-Clusterhead and be resolved within the cooperation cluster.This alternate order will terminate the advantage of the home cluster discovery, which is meant to avoid or decrease time delay by communications with the server and the dependence on expensive pathways. If the home cluster discovery is placed before the historical-based discovery, it means that a request is first forwarded to the D-Clusterhead in the cooperation cluster zone regardless whether this request has been passed before and then checks who else has forwarded the same data request and send requests to the previous requester. This ordering terminates the meaning of historical-based discovery, which is able to resolve a data requests with minimum cost of latency and to improve request success ratio.To decrease the bandwidth consumption and data request latency, the number of hops between the data cache and the requester should be as small as possible. Most mobile nodes, however, do not have sufficient cache space, and therefore the caching strategy is to be devised efficiently. The details description of this scheme is illustrated in the next points, while the Algorithm in Figure 4illustrates a formal description of the proposed cache discovery scheme.

*- Local Cache Discovery*

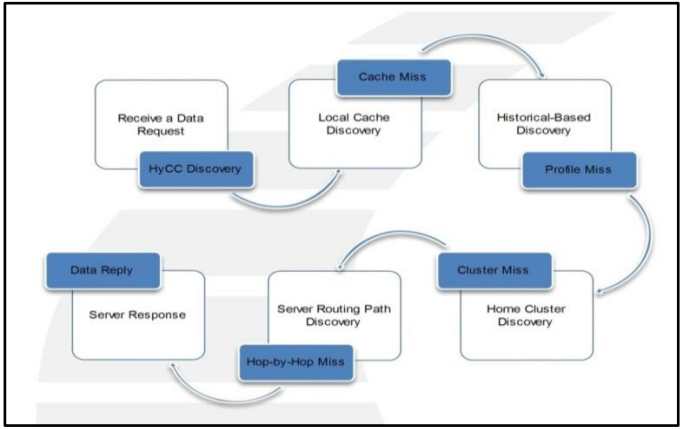

Figure 3. The order of the cache discovery scheme (the proposed model)

When a data request dx is initiated at a node MNi, it is first looks for the data item in its own cache, when a valid copy of the requested data item dx is stored inside the local cache of the requester, the data item is retrieved to serve the request of node application and no cooperative caching is necessary.

*- Historical-Based Discovery*

When the requested item dx is not stored in the local cache i.e. cache miss, as mentioned before, the system called a new strategy for data cache discovery which depends on a historical-based scheme. It maintains a historical file of previously received data requests called PreReq table, each entry records the specific information about previously received request which passed by the node before.

In operation, a node checks PreReq after its local cache misses. If matching entries are found, the node compares its network distance to the newest matching caches and the original data server, and then selects the closest one to send the data request. If the requested data is successfully returned, it means the PreReq entry is still good. Otherwise if the data request fails, it means the cache recorded in the PreReq entry does not have the requested data. Consequently, the PreReq entry is removed; it is the last chance for the requesting node MNi to get any information about the requested data item dx

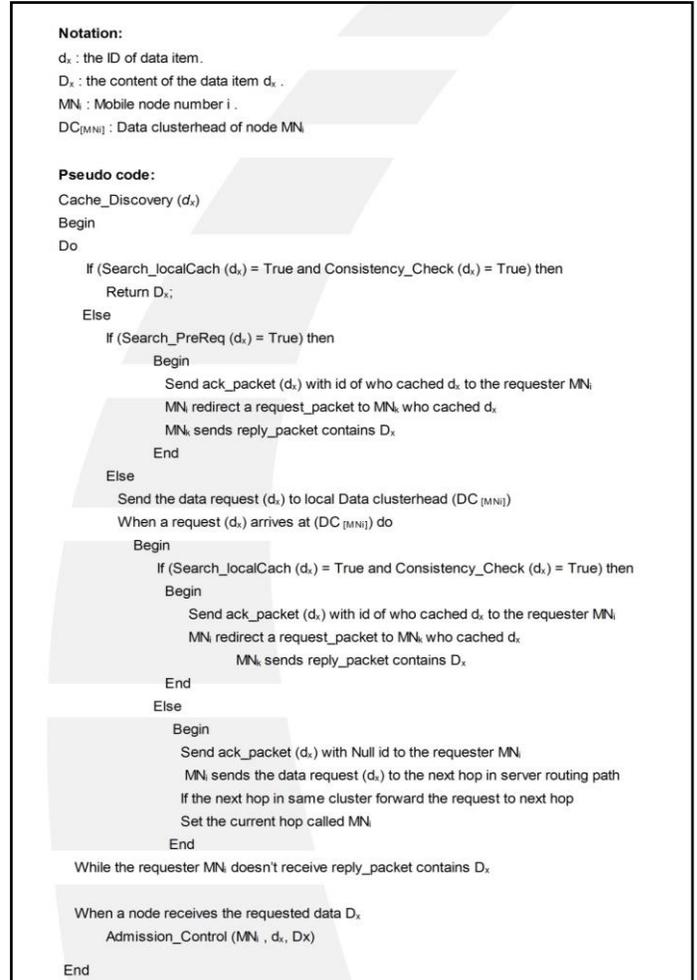

from the node itself without broadcasting or server fetching.

Figure 4. Algorithm of cache discovery scheme

*- Home Cluster Discovery*

To extend the cooperation between Ad-hoc mobile nodes, this paper came up with the home cluster discovery scheme. As mentioned in pervious sections, a typical cluster consists of a D-Clusterhead and a number of nodes, with each node belonging to one cluster. If the nodeMNi cannot get the requested data dx from its local cache or by using any location information in its PreReq table i.e. local miss, the node will search for the requested data dxfrom the members of its cluster by sending a lookup massage packet to the D-Clusterhead in its cluster. When it receives the lookup massage, the D-Clusterhead searches its cache for a valid copy of the requested data item dx. If the item dxis found, the D-Clusterhead will send an acknowledging massage packet that it has the requested data item Dx, or the ID of the node who has cached the requested data item. When a requesting node receives this acknowledgement, it sends a confirming message packet to the node that has the data and that node responds with a reply message that contains the requested data item Dx. Otherwise, if no node has the data item the D-Clusterhead will send a massage packet to the requester that contains no ID.

*- Server Routing Path Discovery*

In case of a home cluster miss, the request is forwarded to neighbor, i.e. the next hop along the routing path. Before forwarding the request of a data item from a node, each intermediate node will check its local cache. If it has the data, it will send it directly and stop the forwarding. Otherwise it will check its PreReq table; if it contains matching entries, it will reply an acknowledging massage packet that it has the data item dx containing the ID of the node who has cached the requested data item. Otherwise forward the request to the D-Clusterhead as described above.

*- Access Data Server*

If the requested data item is not found on the local cache or clusters along the routing path, the data request will finally reach the data server, and the data item dx is retrieved from the data server.

## VI. CONCLUSION

This paper has designed hybrid cooperative cache discovery, which improved data access efficiency in MANET. For cache discovery, we developed a proactive approach, depends on record the requests in historical file, and proposed a hybrid discovery technique used clustering philosophy. In a proactive approach, a node has a PreReq table contains six specific fields related to each item record; a node checks it after its local cache misses and before forwards a request.In heterogeneous mobile Ad-hoc networks, some nodes may be more powerful and have more capacity than other nodes in one cluster. The proposed discovery scheme extended by presented election strategy and managing scheme for these powerful nodes, which called D-Clusterhead, which is responsible for maintaining the information of cluster cache state table of different nodes belonging to the cluster boundary, and provide additional service during cache discovery and cache management. The expected results which will be published later present that the hybrid cache discovery algorithm with these two contributions performs better in terms of cache hit ratio and average message in comparison with compared strategies.

As a future work the researchers will start working on the proposed model to simulate and evaluate it by NS2 simulator, and will be motivated by all these issues and come up with the hybrid cooperative caching strategy scheme based on a combination of cache discovery and cache management algorithms which should be flexible, usable, and implemental not depend on any specific routing protocol.